\documentclass[letterpaper, 10 pt, conference]{ieeeconf}  

\IEEEoverridecommandlockouts                              
\usepackage{amsmath}
\usepackage{float}
\usepackage{graphicx}
\usepackage[colorlinks=true, allcolors=blue]{hyperref}
\usepackage{mathrsfs}
\usepackage{multicol}
\usepackage{tikz} 
    \usetikzlibrary {automata,positioning}

\usepackage{amssymb}  

\title{\LARGE \bf
Design, Modelling and Analysis of a \\ Bio-inspired Spiking Temperature Regulator
}

\author{J.M. Rosito, E. Petri, E. Steur, W.P.M.H. Heemels 
\thanks{Jasper Rosito, Elena Petri, Erik Steur and Maurice Heemels are with the Department of Mechanical Engineering,
        Eindhoven University of Technology, 5600 MB Eindhoven, The Netherlands
        {\tt\small j.m.rosito@student.tue.nl}, {\tt\small \{e.petri, e.steur, m.heemels\}@tue.nl}}}%

\begin{document}

\maketitle
\thispagestyle{empty}
\pagestyle{empty}

\addtolength{\textheight}{-3cm}   

\begin{abstract}

In biology, homeostasis is the process of maintaining a stable internal environment, which is crucial for optimal functioning of organisms. One of the key homeostatic mechanisms is thermoregulation that allows the organism to maintain its core temperature within tight bounds despite being exposed to a wide range of varying external temperatures. Instrumental in thermoregulation is the presence of thermosensitive neurons at multiple places throughout the body, including muscles, the spinal cord, and the brain, which provide spiking sensory signals for the core temperature. In response to these signals, thermoeffectors are activated, creating a negative spiking feedback loop. Additionally, a feedforward signal is provided by warmth and cold-sensitive neurons in the skin, offering a measure for the external temperature. This paper presents an electronic circuit-based architecture design to replicate the biological process of thermoregulation, combined with a formal mathematical analysis. The considered architecture consists of four temperature sensitive neurons and a single actuator, configured in a negative feedback loop with feedforward control. To model the overall system mathematically, hybrid dynamical system descriptions are proposed that are used to analyze and simulate the performance of the design. The analysis and numerical case study illustrate the crucial role of feedforward control in reducing the dependency on the external temperature.

\end{abstract}

\section{INTRODUCTION}

\begin{figure}
    \centering
    \includegraphics[width=\linewidth]{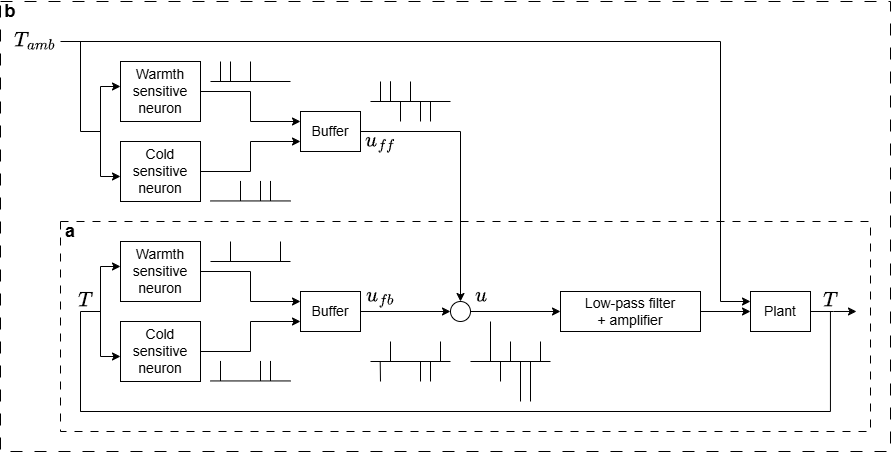}
    \caption{a) Block diagram representing the architecture of a temperature regulator with two temperature sensitive neurons configured in a negative feedback loop as in \cite{original}. b) The newly proposed temperature regulating systems architecture where, in addition to the feedback loop, two additional temperature sensitive neurons are implemented providing a feedforward signal for compensating for fluctuation ambient temperature.}
    \label{fig:blockdiagramfeedforward}
\end{figure}

Homeostasis is the physiological process of keeping the internal environment of organisms constant. For endothermic organisms, one internal variable to keep constant is the body's core temperature. Especially humans regulate their core temperature accurately independent of the surrounding temperature \cite{animaltemp}. Temperature is sensed inside the body by two types of thermosensitive neurons: warmth-sensitive neurons, which increase their firing rate with increasing temperature, and cold-sensitive neurons, which increase their firing rate with decreasing temperature \cite{thermomechanisms}. These neurons send spike signals to the hypothalamus, which reacts by activating the proper thermoeffectors such as constricting/widening blood vessels in the skin, sweating, and shivering. A negative feedback loop is formed by combining the sensing neurons and the thermoeffectors. In addition to the negative feedback, temperature sensitive neurons in the skin are used as a proxy for the ambient temperature, providing the body with a feedforward signal \cite{thermomechanisms}.
Inspired by this mechanism, \cite{original} explored whether it is possible to implement this biological mechanism as an analog electronic device: both on a system theoretical level and an experimental hardware implementation. They presented an analog electronic circuit using two heat sensitive neurons to regulate the temperature of a mass with a single effector, a peltier element. This ``neuromorphic" circuit implements a negative feedback loop which proves promising in an experimental setting. A schematic representation is shown in Fig.~\ref{fig:blockdiagramfeedforward}.a (by the dashed box indicated with `a'). However, in the feedback-only configuration, the equilibrium of the regulated temperature still depends significantly on the ambient temperature, leading to large variations in core temperature, not matching the accurate core temperature regulation observed in biology \cite{animaltemp}. This dependency is observed during experiments and acknowledged by the authors in \cite{original}.

We aim to improve the neuromorphic circuit and controller in \cite{original} to achieve accurate regulation of the core temperature in the presence of fluctuating ambient temperature. To overcome the dependency of the core temperature on the ambient temperature observed in \cite{original} and thus realize a neuromorphic analog device closer to nature, we propose a new architecture design (see Fig.\ref{fig:blockdiagramfeedforward}.b), which augments the architecture of \cite{original} with a feedforward mechanism that uses two additional neurons that sense the ambient temperature. The addition of feedforward control represents the biological mechanism for core temperature regulation in a more realistic manner and minimizes the dependency on the ambient temperature. A circuit implementation of our neuromorphic control architecture will also be presented.

Our second contribution is to show the benefits for design and analysis of modelling these neuromorphic systems in the form of hybrid systems \cite{hybridsystems}. In this paper two of these mathematical models are developed. The first model is a high-fidelity hybrid model of the dynamics of the closed-loop neuromorphic control system. The second model replaces the detailed dynamics of the neurons with integrate-and-fire dynamics, see, e.g., \cite{lapicque1907recherches,abbott1999lapicque, izhikevich2010hybrid}. These models facilitate numerical simulations that turn out to be important for the subsequent analysis of the system, comparing the design in \cite{original} without ambient sensory neurons (Fig.~\ref{fig:blockdiagramfeedforward}.a) and with these neurons (Fig.~\ref{fig:blockdiagramfeedforward}.b). A third model, being an ``averaged" smooth model, is determined to tune the feedforward parameters, but also to demonstrate the obtained performance gain compared to the design in \cite{original} mathematically. Lastly, the results of the tuned design are illustrated in a high-fidelity numerical case study showing the benefits of our novel design.

This paper is organized as follows. In Section~\ref{sec:architecture} the system architecture is described with the modelling of all dynamics of the individual components. These dynamics are formally combined into two hybrid system models in Section~\ref{sec:hybrid}. In Section~\ref{sec:analysis} an analysis of an averaged system model is performed, with and without the feedforward mechanism. Lastly, in Section~\ref{sec:numeric} the results of a numerical case study are presented, and Section~\ref{sec:conclusion} presents conclusions and possible future work.

\section{SYSTEM DESCRIPTION AND ARCHITECTURE}
\label{sec:architecture}

In this section, we propose the overall architecture of the new neuromorphic thermoregulator that includes sensory neurons, coupled to a plant for which the temperature needs to be regulated. A complete circuit layout for the thermoregulator can be found in Fig.~\ref{fig:circuit_ambientsensing}. The plant represents the organisms' body with heating and cooling processes, and is modelled with core temperature $T\in \mathbb{R}$, whose time evolution is described, following \cite{original}, by
\begin{equation}
    \dot{T}=\alpha(T_{amb}-T)-A(V_{out}),
    \label{eq:temperature}
\end{equation}
with ambient temperature $T_{amb}\in \mathbb{R}$, the heat exchange with the surroundings $\alpha>0$, filtered and amplified control signal $V_{out}\in \mathbb{R}$ and actuation function $A(V_{out})$ representing the effect of the thermoeffectors. $V_{out}$ is the result of the spiky signals from temperature-sensitive neurons. $A:\mathbb{R}\rightarrow\mathbb{R}$ is assumed to be monotonically increasing, taking both positive and negative values, and $A(0)=0$.

\begin{figure}
    \centering
    \includegraphics[width=\linewidth]{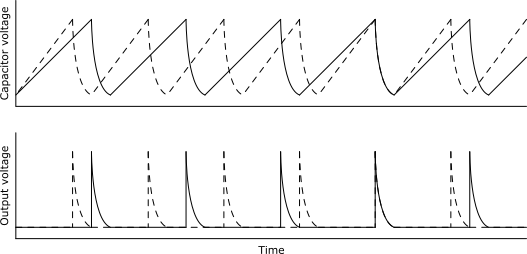}
    \caption{Internal capacitor voltage and output voltage of a neuron over time. The increased slope of the dashed line is induced by a change in temperature, indicating an increase in temperature for a warmth-sensitive neuron or indicating a decrease in temperature for a cold-sensitive neuron.}
    \label{fig:neuronschematic}
\end{figure}

\begin{figure}
    \centering
    \includegraphics[width=.96\linewidth]{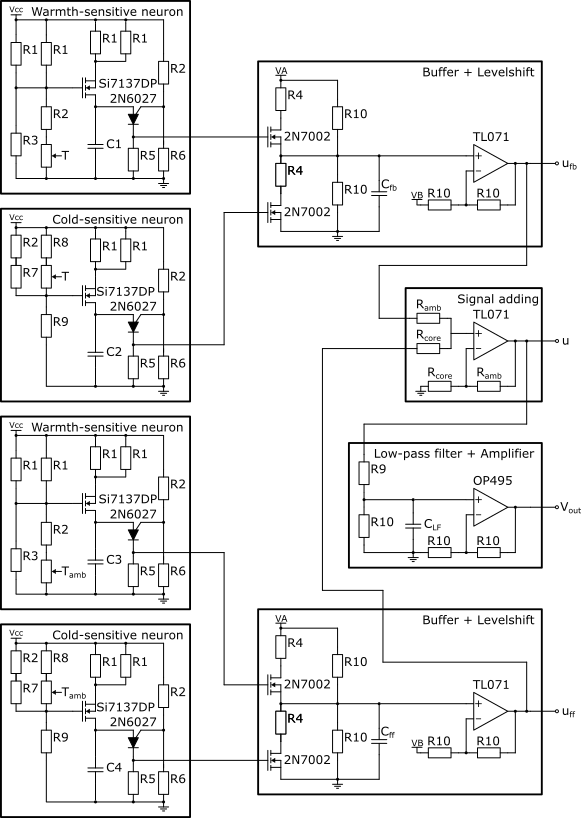}
    \caption{Electronic circuit implementation of the temperature regulating neuromorphic device of Fig~\ref{fig:blockdiagramfeedforward}. With two neurons for sensing the core temperature (top) and two for sensing the ambient temperature (bottom). The signals of both sets are combined in their own respective buffer and then added with their weight set by the two resistors $R_{core}$ and $R_{amb}$. The combined signal is then fed through a low-pas filter and amplified before being sent to the peltier actuator.}
    \label{fig:circuit_ambientsensing}
\end{figure}

The neurons are based on circuits and inspired by the dynamics of integrate-and-fire models, see, e.g., \cite{lapicque1907recherches,abbott1999lapicque, izhikevich2010hybrid}. A capacitor is charged until its voltage $V_i\in \mathbb{R}$ reaches a threshold voltage $V_{on}$, at which a switched is toggled ``on" ($S_{i,out}=1$) and, subsequently, the capacitor is rapidly discharged until $V_i$ hits $V_{off}<V_{on}$. At $V_{off}$, the switch is toggled ``off" ($S_{i,out}=0$) and the capacitor is charged again. The output of the neuron is connected to the ground side of the switch. Therefore, the output voltage is equal to ground when the switch is turned ``off", and equal to the capacitor voltage when the switch is turned ``on" creating a spiking signal. A typical evolution of the capacitor voltage and the output voltage of an electronic neuron over time are shown in Fig.~\ref{fig:neuronschematic}. The neuron circuits include temperature-sensitive resistors, which make the slope in the charging state, and thus the neuron's firing frequency, temperature dependent. For warmth-sensitive neurons, the slope increases for increasing temperature, and for cold-sensitive neurons the slope decreases for increasing temperature. The internal capacitor voltages $V_i\in \mathbb{R}$ and the states of the output switches $S_{i,out}\in\{0,1\}$ with $i\in\{1,2,3,4\}$ are taken as the states of the neurons. The evolution of $V_i$ can be modelled by
\begin{equation}
    \dot{V}_{i}=\frac{1}{C_{i}} \left( I_{FET}(V_{G,i}) -\frac{V_i S_{i,out}}{R_5} \right),
    \label{eq:neuron}
\end{equation}
with its corresponding capacitance $C_i$, the charge current $I_{FET}(V_{G,i})\in \mathbb{R}$ as function of the gate voltage $V_{G,i}$ and the discharge current $\frac{V_i S_{i,out}}{R_5}$, see Fig.~\ref{fig:circuit_ambientsensing} for the context within the circuit. Note that the capacitors discharge when the switch is in its ``on" state, i.e., $S_{i,out}=1$, and charge when the switch is in its ``off" state, i.e., $S_{i,out}=0$. The switch mechanism described above for $S_{i,out}$ will be formalized in the models in the next section.

The gate voltages $V_{G,i}$ of the MOSFETs of warmth-sensitive neurons are given by
\begin{equation}
    V_{G,1}=\frac{V_{cc}}{1+\frac{1}{2}R_1 \left( R_3^{-1}+\left(R_{NTC}(T)+R_2\right)^{-1}\right)},
    \label{eq:gate1}
\end{equation}
\begin{equation}
    V_{G,3}=\frac{V_{cc}}{1+\frac{1}{2}R_1 \left( R_3^{-1}+\left(R_{NTC}
    (T_{amb})+R_2\right)^{-1}\right)},
    \label{eq:gate3}
\end{equation}
where $R_j$, $j\in\{1,2,3,...,10\}$ denote relevant resistances in the circuit architecture (see Fig.~\ref{fig:circuit_ambientsensing}), $V_{cc}$ denotes the constant circuits supply voltage, and $R_{NTC}(T)$ is the temperature-dependent resistance . $R_{NTC}$ is strictly decreasing with strictly increasing temperature \cite{NTCLE100E3}. Similarly, the gate voltages of the MOSFETs of cold-sensitive neurons are given by
\begin{equation}
    V_{G,2}=\frac{V_{cc}R_9}{R_9+\left( \left( R_2+R_7 \right)^{-1} + \left(R_8+R_{NTC}(T)\right)^{-1}\right)^{-1} }
    \label{eq:gate2}
\end{equation}
\begin{equation}
    V_{G,4}=\frac{V_{cc}R_9}{R_9+\left( \left( R_2+R_7 \right)^{-1} + \left(R_8+R_{NTC}(T_{amb})\right)^{-1}\right)^{-1} }.
    \label{eq:gate4}
\end{equation}
Note that for (\ref{eq:gate1}) and (\ref{eq:gate3}) the gate voltages decrease for increasing temperature and for (\ref{eq:gate2}) and (\ref{eq:gate4}) the gate voltages increase for increasing temperature.

For a reasonable temperature range ($0-100^{\circ}C$), it can be assumed that the gate voltages are such that the MOSFETs operate in their saturation region \cite{Si7137DP}. In this regime, the drain current depends only on the gate-source voltage. Since the current is very small, the voltage drop over the resistors at the source can be neglected, and the source voltage can be taken as $V_S=V_{cc}$. With the source voltage fixed, the drain current can be given as function of $V_{G,i}$ by
\begin{equation}
    I_{FET}(V_{G,i})=K_p \left(V_{G,i}-V_S-V_{th} \right)^2
    \label{eq:ifet}
\end{equation}
with transconductance $K_p$ and threshold voltage $V_{th}$, which are physical parameters of the MOSFET. For a MOSFET in its saturation region, the term $V_{G,i}-V_S-V_{th}<0$ in \eqref{eq:ifet} therefore, the charge current $I_{FET}$ increases with decreasing gate voltages, as is the case for the warmth-sensitive neurons (see (\ref{eq:gate1}) and (\ref{eq:gate2})). Resulting in neurons (defined in (\ref{eq:neuron})) that increase their firing frequency when the temperature increases. Similarly, the increasing gate voltages of the cold-sensitive neurons in (\ref{eq:gate3}) and (\ref{eq:gate4}), result in neurons with decreasing firing frequency with increasing temperature.

The output signals of the neurons are the voltages on the ground side of the output switches and are given by
\begin{equation}
    V_{i,out}=V_i S_{i,out}
\end{equation}

The output signals of the neurons are used to control the internal capacitor voltage of a buffer via input switches, see Fig.~\ref{fig:blockdiagramfeedforward}. The input switches of the buffer, denoted as $S_{i,b}$ with $i\in\{1,2,3,4\}$, electrically isolate the neurons from the buffer, so that no current flows between them. When an input switch corresponding to a cold-sensitive neuron is ``on" ($S_{i,b}=1$ with $i\in\{2,4\}$), the buffer has an exponentially stable equilibrium at $0V$. During a spike, the voltage flows for a short amount of time in that direction. Similarly, when an input switch corresponding to a warmth-sensitive neuron is ``on" ($S_{i,b}=1$ with $i\in\{1,3\}$), the voltage has an exponentially stable equilibrium at its upper limit $V_A$. During a spike, the voltage flows for a short amount of time in that direction. If no input switches are ``on", i.e., $S_{i,b}=0$ for all $i\in\{1,2,3,4\}$, a small leakage current provides an exponentially stable equilibrium at $\frac{1}{2}V_A$. The difference in the firing frequency of the warmth and cold-sensitive neuron dictates the voltage around which the capacitor finds a new equilibrium voltage. Two separate buffers are used for the core temperature-sensing neurons and for the ambient temperature-sensing neurons, providing a feedback signal, related to neurons $i \in \{1,2\}$, and feedforward signal, corresponding to neurons $i \in \{3,4\}$. The buffers have their internal capacitor voltage $V_{fb}\in \mathbb{R}$ and $V_{ff}\in \mathbb{R}$, the warmth-sensitive neuron input switches $S_{1,b}\in\{0,1\}$ and $S_{3,b}\in\{0,1\}$, and the cold-sensitive neuron input switches $S_{2,b}\in\{0,1\}$ and $S_{4,b} \in \{0,1\}$ as their states. The evolution of the internal capacitor voltages can be modelled by
\begin{equation}
    \begin{split}
        \dot{V}_{fb}=-\frac{2V_{fb}-V_A}{C_{fb} R_{10}}+\frac{V_A-V_{fb}}{C_{fb} R_4}S_{1,b} -\frac{V_{fb}}{C_{fb} R_4}S_{2,b},
    \end{split}
    \label{eq:bufferfb}
\end{equation}
\begin{equation}
    \begin{split}
        \dot{V}_{ff}=-\frac{2V_{ff}-V_A}{C_{ff} R_{10}}+\frac{V_A-V_{ff}}{C_{ff} R_4}S_{3,b} -\frac{V_{ff}}{C_{ff} R_4}S_{4,b}.
    \end{split}
    \label{eq:bufferff}
\end{equation}
The state of the input switches is ``on", i.e., $S_{i,b}=1$, when the output voltage of the corresponding neuron is above a threshold voltage $V_{th}<V_{on}$ and ``off", i.e., $S_{i,b}=0$, otherwise. The first term in (\ref{eq:bufferfb}) and (\ref{eq:bufferff}) provides the leakage current for the neutral position $\frac{1}{2}V_A$. The buffer voltage is also at its neutral position, if the firing frequency of both corresponding neurons is equal, i.e., their charge current and, therefore, gate voltages are equal, (\ref{eq:gate1}) and (\ref{eq:gate2}) for the feedback buffer and (\ref{eq:gate3}) and (\ref{eq:gate4}) for the feedforward buffer. The temperature at which this is the case is defined as the system-inherent setpoint $T_{set}\in \mathbb{R}$.
The internal capacitor voltages of the buffers are shifted to place the neutral position at $0V$. The resulting voltages are the control signals given by $u_j=V_j-\frac{1}{2}V_A$, $j\in\{fb,ff\}$. Both control signals are combined into one signal given by
\begin{equation}
    u=u_{fb}+Ku_{ff},
\end{equation}
where $K>0$ is the gain of the feedforward signal.

The control signal $u$ is fed through a low-pass filter, see Fig.~\ref{fig:blockdiagramfeedforward}, with capacitance $C_{LP}$, and the capacitor voltage $V_{LP}\in\mathbb{R}$ evolves according to
\begin{equation}
    \dot{V}_{LP}=\frac{1}{C_{LP}R_9}(u-V_{LP})-\frac{1}{C_{LP}R_{10}}V_{LP}.
    \label{eq:lowpass}
\end{equation}
This filtered signal is then amplified to create the output voltage 
\begin{equation}
    V_{out}=\left(1+\frac{R_4}{R_5} \right)V_{LP},
    \label{eq:vout}
\end{equation}
going into (\ref{eq:temperature}). This provides the complete system description and architecture design we propose.

In Section~\ref{sec:hybrid} we provide formal hybrid systems models for this setup.

\section{HYBRID MODELLING}
\label{sec:hybrid}

In this section, the architecture in Section~\ref{sec:architecture} will be captured into two hybrid models using the formalism of \cite{hybridsystems}. The first model implements the dynamics of the proposed electronic circuit, as described in the previous section, in a framework that can be simulated, fully modeling the spikes as fast continuous signals, as in Fig.~\ref{fig:neuronschematic}. This model is then used as a bridge to the second hybrid model in which the spikes are considered as instantaneous pulses, essentially assuming that the time constants of the spikes are significantly smaller than those of the remaining dynamics, which is reasonable. This allows for a model with fewer states that can be simulated more efficiently.

\subsection{Hybrid Model A: Spikes as fast continuous signals}

A (physical) state vector $x\in \mathbb{R}^{n_x}$ with $n_x=8$ is defined consisting the temperature and all relevant voltages, and is given by
\begin{equation}
    x:=\left( T,V_1,V_2,V_3,V_4,V_{fb},V_{ff},V_{LP} \right)^T
    \label{eq:x}
\end{equation}
In addition, a state vector $S\in\{0,1\}^{n_S}$, with $n_S=8$, consisting of the states of all switches is defined by
\begin{equation}
    \begin{split}
        S= ( S_{1,out},S_{2,out},S_{3,out},S_{4,out},S_{1,b},S_{2,b},S_{3,b},S_{4,b} )^T .
    \end{split}
\end{equation}
The overall state vector is defined as $q:=(x,S)\in \mathcal{Q}$ with $\mathcal{Q}:=\mathbb{R}^{n_x} \times \{0,1\}^{n_S}$ used in the hybrid system setup
\begin{equation}
    \begin{cases}
      \dot{q}=F(q), & q\in C,\\
      q^+\in G(q), & q\in D.
    \end{cases} 
    \label{eq:systemA}
\end{equation}
Here, $C\subseteq \mathcal{Q}$ is the flow set, $D\subseteq \mathcal{Q}$ is the jump set, $F:\mathcal{Q}\rightarrow\mathcal{Q}$ is the flow map and $G:\mathcal{Q}\rightarrow\mathcal{Q}$ is the jump map, see \cite{hybridsystems} for more details on this hybrid modelling framework.
The flow map $F$, for any $q\in C$, is then
\begin{equation}
    F(q):=\left[f(q),0_{n_S}^T\right]^T,
\end{equation}
where $f(q)$ can easily be derived from (\ref{eq:temperature})-(\ref{eq:vout}), and $0_{n_S}$ represents the non-flowing nature of the discrete states in $S$.

The flow set $C$ is defined as
\begin{equation}
    \begin{split}
        C:=&C_{1,out} \cap C_{2,out} \cap C_{3,out} \cap C_{4,out}... \\
        &\cap  C_{1,b} \cap C_{2,b} \cap C_{3,b} \cap C_{4,b}
    \end{split}
\end{equation}
with
\begin{equation}
    \begin{split}
        &C_{i,out}:=\{q\in \mathcal{Q}|(S_{i,out}=0 \land V_i\leq V_{on}) \\
        &\ \lor (S_{i,out}=1\land V_i\geq V_{off}) \}, i\in\{1,2,3,4\},
    \end{split}
    \label{eq:CSjumpneuron}
\end{equation}
\begin{equation}
    \begin{split}
        &C_{i,b}:=\{q\in \mathcal{Q}|(S_{i,b}=0\land V_{i,out}\leq V_{th}) \\
        &\ \lor (S_{i,b}=1\land V_{i,out)}\geq V_{th}) \}, i\in\{1,2,3,4\}.
    \end{split}
\end{equation}

The jump set $D$ is defined as
\begin{equation}
    \begin{split}
        D:=&D_{1,out}\cup D_{2,out}\cup D_{3,out}\cup D_{4,out}... \\
        &\cup D_{1,b}\cup D_{2,b}\cup D_{3,b}\cup D_{4,b}.
    \end{split}
\end{equation}
with
\begin{equation}
    \begin{split}
        &D_{i,out}:=\{q\in \mathcal{Q}|(S_{i,out}=0 \land V_i\geq V_{on}) \\
        &\ \lor (S_{i,out}=1\land V_i\leq V_{off}) \}, i\in\{1,2,3,4 \},
    \end{split}
    \label{eq:CSjumpneuron}
\end{equation}
\begin{equation}
    \begin{split}
        D_{i,b}:=\{q\in \mathcal{Q}|(S_{i,b}=1\land V_{i,out}\leq V_{th}) \\
        \lor (S_{i,b}=0\land V_{i,out)}\geq V_{th}) \}, i\in\{1,2,3,4\}.
    \end{split}
\end{equation}

The jump map $G$ is defined as
\begin{equation}
    \begin{aligned}
        G(q):=&G_{1,out}(q)\cup G_{2,out}(q)\cup G_{3,out}(q)
        \cup G_{4,out}(q)... \\ 
        &\cup G_{1,b}(q)\cup G_{2,b}(q)\cup G_{3,b}(q)\cup G_{4,b}(q)
        \end{aligned}
\end{equation}
with
\begin{equation}
    G_{i,out}(q):=\begin{cases}
        \begin{matrix}
            \begin{pmatrix}
                x \\
                \Lambda_iS_{out}+\Gamma_i \\
                S_b
            \end{pmatrix} & q\in D_{i,out} \\
            \emptyset & q\notin D_{i,out},
        \end{matrix}
    \end{cases} 
\end{equation}
\begin{equation}
    G_{i,b}(q):=\begin{cases}
        \begin{matrix}
            \begin{pmatrix}
                x \\
                S_{out} \\
                \Lambda_iS_b+\Gamma_i \\
            \end{pmatrix} & q\in D_{i,b} \\
            \emptyset & q\notin D_{i,b},
        \end{matrix}
    \end{cases} 
    \label{eq:GfbH}
\end{equation}
where $\Lambda_i\in\mathbb{R}^{4\times 4}$ is the diagonal matrix with all elements on the diagonal being equal to 1 except for the $i$-th diagonal element, which is -1, $S_{out}:=(S_{1,out},S_{2,out},S_{3,out},S_{4,out})^T$ is a vector containing all the current states of the output switches, $\Gamma_i\in \mathbb{R}^4$ is a vector with all elements equal to 0 except the $i$-th element, which is equal to 1, $S_b:=(S_{1,b},S_{2,b},S_{3,b},S_{4,b})^T$ is a vector containing all the current states of the buffer input switches, and $\emptyset$ denotes the empty set.

\subsection{Hybrid Model B: Spikes as instantaneous pulses}
\label{sec:modelB}
The second model describes the fast dynamics of the spikes as instantaneous pulses. As a result, switch states are omitted from the overall state vector, now defined as $\bar{q}:=x$, note $x$ is as defined in (\ref{eq:x}), as they are only active during spikes in Fig.~\ref{fig:neuronschematic}, which are now made instantaneous. This leads to hybrid system
\begin{equation}
    \begin{cases}
      \dot{\bar{q}}=\bar{F}(\bar{q}), & \bar{q}\in \bar{C},\\
      \bar{q}^+\in \bar{G}(\bar{q}), & \bar{q}\in \bar{D},
    \end{cases} 
    \label{eq:systemB}
\end{equation}
where we now have $\bar{C}\subseteq \mathbb{R}^{n_x}$ as the flow set, $\bar{D}\subseteq \mathbb{R}^{n_x}$ as the jump set, $\bar{F}:\mathbb{R}^{n_x} \rightarrow \mathbb{R}^{n_x}$ as the flow map and $\bar{G}:\mathbb{R}^{n_x} \rightarrow \mathbb{R}^{n_x}$ as the jump map.
The flow map $\bar{F}$, for any $\bar{q}\in \bar{C}$, is then defined as
    $\bar{F}(\bar{q}):=\bar{f}(\bar{q})$,
where $\bar{f}(\bar{q})=f(\bar{q},0_{n_S})$.

The flow set $\bar{C}$ is defined as
\begin{equation}
    \bar{C}:=\bar{C}_1\cap \bar{C}_2\cap \bar{C}_3\cap \bar{C}_4
\end{equation}
with
\begin{equation}
    \bar{C}_i:=\{\bar{q}\in \mathbb{R}^{n_x}|V_{i}\leq V_{on}\},\quad i\in \{1,2,3,4\}.
\end{equation}

The jump set $\bar{D}$ is defined as
\begin{equation}
    \bar{D}:=\bar{D}_1\cup \bar{D}_2\cup \bar{D}_3\cup \bar{D}_4
\end{equation}
with
\begin{equation}
    \bar{D}_i:=\{\bar{q}\in \mathbb{R}^{n_x}|V_{i}\geq V_{on}\},\quad i\in \{1,2,3,4\}.
\end{equation}

The jump map $\bar{G}$ is defined by
\begin{equation}
    \bar{G}(\bar{q}):=\bar{G}_1(\bar{q})\cup \bar{G}_2(\bar{q})\cup \bar{G}_3(\bar{q})\cup \bar{G}_4(\bar{q})
\end{equation}
with
\begin{equation}
    \bar{G}_i(\bar{q}):=\begin{cases}
        \begin{matrix}
            \begin{pmatrix}
                T \\
                (I-\Gamma_i\Gamma_i^T)V+\Gamma_i V_{off} \\
                V_b(V_{fb},V_{ff},i) \\
                V_{LP}
            \end{pmatrix} & \bar{q}\in \bar{D}_i \\
            \emptyset & \bar{q}\notin \bar{D}_i,
        \end{matrix}
    \end{cases} 
\end{equation}
where $\Gamma_i\in\mathbb{R}^4$ is as defined after (\ref{eq:GfbH}), $V:=(V_1,V_2,V_3,V_4)^T$, and function $V_b(V_{fb},V_{ff},i)$, with $i\in\{1,2,3,4\}$, gives the buffer voltages after a jump as a function of the current buffer voltages, replacing the fast spiky dynamics. The function depends on which neuron spikes. One function for both buffers is provided, where for the core temperature sensing neurons, i.e., $i\in\{1,2 \}$, $V_{fb}$ jumps and $V_{ff}$ is left unmodified and for the ambient temperature sensing neurons, i.e., $i\in\{3,4 \}$, $V_{ff}$ jumps and $V_{fb}$ is left unmodified. The function is given by
\begin{equation}
    V_b\left(V_{fb},V_{ff},i\right)=\begin{cases}
        \begin{matrix}
            \begin{pmatrix}
                a(\tau)V_{fb}+\left(1-a(\tau)\right)V_A \\
                V_{ff}
            \end{pmatrix}, & i=1 \\
            \begin{pmatrix}
                a(\tau)V_{fb} \\
                V_{ff}
            \end{pmatrix}, & i=2 \\
            \begin{pmatrix}
                V_{fb} \\
                a(\tau)V_{ff} +\left(1-a(\tau)\right)V_A
            \end{pmatrix}, & i=3 \\
            \begin{pmatrix}
                V_{fb}\\
                a(\tau)V_{ff}
            \end{pmatrix}, & i=4
        \end{matrix}
    \end{cases}
\end{equation}
where $a(\tau)$ is the jump ratio, which is obtained by solving the differential equations (\ref{eq:bufferfb}) and (\ref{eq:bufferff}) during the discharge of the neuron's capacitor voltage $V_i$, $i \in \{1,2,3,4\}$, which has duration $\tau$ and neglecting the leakage current. Spike duration $\tau$ is derived similarly by solving  the (\ref{eq:neuron}) during discharging of the neuron, leading to
\begin{equation}
    a(\tau)=e^{-\frac{1}{C_{fb} R_4}\tau},
\end{equation}
with
\begin{equation}
    \tau = -C_iR_5ln\left(\frac{V_{off}}{V_{on}}\right).
\end{equation}

In the next section, an ``averaged" smooth model is obtained by simulating the hybrid systems described in this section. This averaged model is used to analyze and tune the feedforward gain and demonstrate the improved performance when implementing ambient temperature sensing neurons (feedforward).

\section{Averaged model and analysis}
\label{sec:analysis}
To show the benefits of adding feedforward control, a model of the system architecture with and without the ambient temperature sensing neurons is analyzed.

\subsection{The case without feedforward}
An averaged model with only two neurons in negative feedback loop (as in Fig.~\ref{fig:blockdiagramfeedforward}.a and following \cite{original}) can be obtained as
\begin{subequations}
    \begin{align}
        \dot{T}&=\alpha \left(T_{amb}-T \right)-B(u), \label{eq:T} \\
        \dot{u}&=-\gamma u +r_H-r_C. \label{eq:u}
    \end{align}
\end{subequations}
with core temperature $T\in\mathbb{R}$, ambient temperature $T_{amb}\in\mathbb{R}$, the input signal of the actuator $u\in\mathbb{R}$ and functions $r_H>0$ and $r_C>0$ representing the spike trains of the warmth-sensitive and cold-sensitive neuron on the buffer voltage. The quantities $\gamma>0$ and $\alpha>0$ are system parameters and $B(u)$ represents the actuator function. When relating this model to the system architecture described in Section~\ref{sec:architecture}, $\alpha$ is the heat exchange coefficient with the surroundings as in (\ref{eq:temperature}), $u$ is equal to the feedback control signal (as we consider no feedforward, as in \cite{original}), $\gamma$ is the buffers leakage term in (\ref{eq:bufferfb}), and mapping $B$ captures the combined properties of $A$, the low-pass filter and the output amplifier. Assuming that the dynamics of the spikes and the dynamics of $u$ are significantly faster than the dynamics of $T$, a new function $\tilde{u}(T)$ can be defined, which represents the equilibrium (averaged) value $\tilde{u}(T)$, where the dynamics of (\ref{eq:u}) converge to. This depends only on the temperature. This replaces the dynamics of (\ref{eq:T}) by
\begin{equation}
    \dot{T}=-\alpha \left(T_{amb}-T \right)-B(\tilde{u}(T))
    \label{eq:simpleT}
\end{equation}
The shape of function $\tilde{u}$ is determined by simulations (assuming time-scale separation), based on the hybrid systems models in Section~\ref{sec:hybrid}; we describe this in Section~\ref{sec:numeric} below. As shown in Fig.~\ref{fig:u_Vs_T}, $\tilde{u}$ has typically sigmoidal shape. The temperature $T$ at which $\tilde{u}(T)=0$ is considered the system-inherent set point and is denoted by $T_{set}\in\mathbb{R}$. For the $\tilde{u}$ in Fig.~\ref{fig:u_Vs_T}, $T_{set}\approx 39,84^{\circ}C$. This is related to the system-inherent setpoint described in Section~\ref{sec:architecture}, which follows from the feedback buffer dynamics defined in (\ref{eq:bufferfb}). With the defined set point $T_{set}$, new variables for the error and the disturbance are defined by $e(t):=T(t)-T_{set}$ and $d(t):=T_{amb}(t)-T_{set}$. Also, a new function for the actuation is defined as $\tilde{B}(e)=B(\tilde{u}(e+T_{set}))$. The error dynamics are then given by
\begin{equation}
    \dot{e}=-\alpha e+\alpha d -\tilde{B}(e).
\end{equation}
Fig.~\ref{fig:u_Vs_T} shows $\tilde{u}$ to be roughly linear for $T\in \left[30^{\circ}C,50^{\circ}C \right]$, corresponding to $\tilde{B}(e)\approx ce$ for $e\in \left[-10,10 \right]$. Using this assumption the error has an exponentially stable equilibrium at 
\begin{equation}
    e^*=\frac{\alpha}{\alpha+c}d,
\end{equation}
where a clear dependency on the ambient temperature, through the presences of $d$, is visible.

\subsection{The case with feedforward}
\label{sec:analysisff}
Adding feedforward control to the model, as proposed in our new system architecture described in Section~\ref{sec:architecture}, a new equation for the error dynamics is given by
\begin{equation}
    \dot{e}=-\alpha e+\alpha d -\tilde{B}(e)-K\tilde{B}(d),
\end{equation}
where $K$ is the gain of the feedforward control signal and $\tilde{B}(d)$ is the averaged feedforward signal related to $u_{ff}$ in Section~\ref{sec:architecture}, caused by the spike trains of the ambient sensory neurons. Since the neurons and buffer circuit parameters for the ambient temperature sensing are chosen identically to their core temperature sensing counterparts, the same actuation function $\tilde{B}$ can be used. For our design the error dynamics has an exponentially stable equilibrium at
\begin{equation}
    e^*=\frac{\alpha-Kc}{\alpha+c}d.
\end{equation}
The influence of the ambient temperature on the equilibrium can be eliminated when taking $K=\frac{\alpha}{c}$, thereby showing explicitly the benefit and importance of the introduction of the ambient sensory neurons (feedforward), realizing accurate temperature regulation over a large range of ambient temperatures. This is closer to what is observed in real endothermic organisms, see the biological study \cite{animaltemp}. We will demonstrate this further in a numerical case study in the next section.

\section{Numerical case study}
\label{sec:numeric}
\begin{figure}
    \centering
    \includegraphics[width=\linewidth]{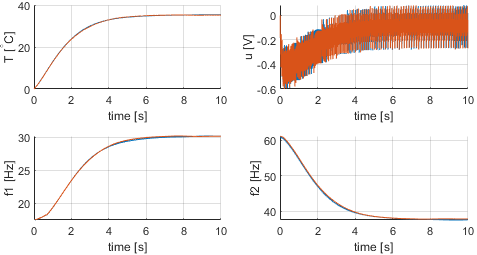}
    \caption{Simulation of both hybrid models without ambient temperature sensing neurons. Hybrid system A in blue and hybrid system B in orange. Core temperature $T [^{\circ}C]$ (top left), input signal $u [V]$ (top right), spike frequency of the hot neuron $f_1 [Hz]$ (bottom left) and spike frequency of the cold neuron $f_2 [Hz]$ (bottom right) are shown.}
    \label{fig:combientmodels}
\end{figure}

The simulations are performed with the Hybrid Equations Toolbox in Matlab \cite{hybridtoolbox}. In the simulations of systems (\ref{eq:systemA}) and (\ref{eq:systemB}) with the dynamics described in Section~\ref{sec:architecture} the components are modelled with the values in Table~\ref{tab:param}, the actuation function is taken as $A(V_{out})=2V_{out}$, $\alpha=2$, $V_{on}=7.4$, $V_{off}=1$, $K_p=5\cdot10^{-6}$, and $V_{th}=0.7$. Fig.~\ref{fig:combientmodels} shows the results of simulating both models with no feedforward control, i.e., $K=0$. Here the blue line is hybrid model A and the orange line is hybrid model B. The figure shows that both models are very close to each other, thereby showing that the spike dynamics in Fig.~\ref{fig:neuronschematic} are indeed significantly faster then that of the other dynamics. For hybrid model A the run time was 11,9s and 1405 jumps were made and for hybrid model B the run time was 2,2s and 705 jumps were made. This clearly shows the benefit of the simplified system B when using simulations to explore the parameter space. The following figures all show results obtained from simulating hybrid system in Section \ref{sec:modelB}.

\begin{figure}
    \centering
    \includegraphics[width=0.9\linewidth]{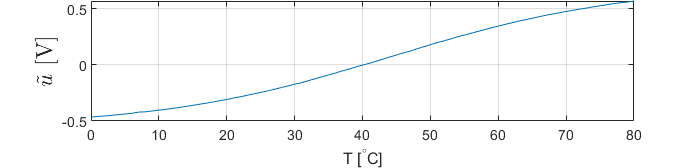}
    \caption{Average value of input signal $\tilde{u}(T)$ as result of $r_H(t)$ and $r_C(t)$. The results are obtained using hybrid model B (see Section~\ref{sec:modelB}) with temperature steps of $1^{\circ}C$ and a period of $20s$ (to assure that $u$ in (\ref{eq:u}) converges (on average) to its steady-state value $\tilde{u}(T)$). Note the zero crossing is at $39,84^{\circ}C$ which is considered $T_{set}$.}
    \label{fig:u_Vs_T}
\end{figure}

The averaged input function $\tilde{u}(T)$ described in Section~\ref{sec:analysis}, see (\ref{eq:simpleT}), can be determined during simulations by measuring $u(t)$ for 20s at a constant temperature and taking its average value. Fig.~\ref{fig:u_Vs_T} shows the resulting sigmoidal function. The temperature where $\tilde{u}(T)=0$ is considered the setpoint $T_{set}=39.84^{\circ}C$, as indicated before.

\begin{figure}
    \centering
    \includegraphics[width=0.97\linewidth]{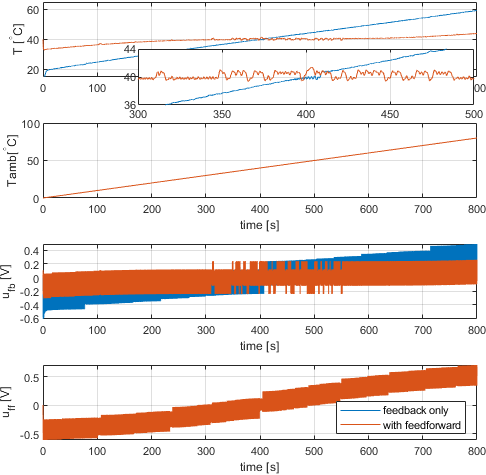}
    \caption{Simulation results of the system without feedforward (blue) and with feedforward (orange). The ambient temperature is slowly increased over 800 seconds to show the dependence of the temperature $T$ on the ambient temperature $T_{amb}$. The zoom shows how the temperature $T$ is constant over the (approximately) linear region of the input function $\tilde{u}(T_n)$ when the ambient signal gain is set at $K=\frac{\alpha}{c}$.}
    \label{fig:experiments}
\end{figure}

The system without and with feedforward are simulated for a duration of 800 seconds. The ambient temperature $T_{amb}$ is slowly increased from $0-80^{\circ}C$ over this period. Fig.~\ref{fig:experiments} shows the results of these simulations. The system without ambient sensing (blue lines) shows a strong dependency on the ambient temperature, as was also shown by our analysis in Section~\ref{sec:analysis}. This does not comply with what is observed in nature \cite{animaltemp}. The system with feedforward (orange lines) shows a significant performance improvement. The feedforward gain is taken as proposed in Section~\ref{sec:analysisff}, as $K=\frac{\alpha}{c}=0.9$. The zoom shows how for a large range of ambient temperatures $T_{amb}\in \left[30^{\circ}C,50^{\circ}C \right]$, the core temperature $T$ is stable close to the setpoint $T_{set}=39.84^{\circ}C$. It can be seen that for this range the feedback control signal $u_{fb}\approx 0$ since $T\approx T_{set}$ and the feedforward control signal $u_{ff}$ is roughly proportional to $T_{amb}-T_{set}$. These results are closer to the accurate homeostasis observed in organisms (see Fig. 1 in \cite{animaltemp}) underlining the importance of including feedforward sensory neurons. Outside this range, the core temperature $T$ starts to deviate from the setpoint $T_{set}$. However, the temperature stays relatively close to the setpoint, despite these extra large variations.

\section{CONCLUSIONS AND FUTURE WORKS}
\label{sec:conclusion}

In this paper, we presented a new electronic circuit-based architecture for recreating biological thermoregulation mechanisms as seen in endothermic animals \cite{animaltemp}. Temperature sensitive ``spiky" electronic neurons and a single thermoeffector are configured in a controller with negative feedback and feedforward control. Two mathematical models in the form of hybrid systems were presented to describe and simulate the system, These models were important for obtaining an ``average" model for analysis. We showed through the analysis and numerical simulations that feedforward is a critical ingredient in minimizing the dependency of the regulated temperature on the ambient temperature. This accurate thermoregulation over large ranges of ambient temperature is close to what is observed in nature \cite{animaltemp}, underpinning the important role of the (feedforward) ambient temperature sensing neurons. The overall design of the bio-inspired thermoregulation system in this paper fits well in the current trend of increasing interest in neuromorphic event-based control, see \cite{sepulchre1}, \cite{petri1}, \cite{sepulchre2} and the references therein, and can bring new insights to this challenging field.

\begin{table}
    \caption{Parameters used for the components in Fig~\ref{fig:circuit_ambientsensing}}
    \label{tab:param}
    \begin{center}
    \begin{tabular}{|c|c|c|c|}
    \hline
        Component & Value & Component & Value \\\hline
        $R_1$ & $39k\Omega$ & $R_{10}$ & $10M\Omega$ \\\hline
        $R_2$ & $100k\Omega$ & $C_1$ & $0.047\mu F$ \\\hline
        $R_3$ & $470k\Omega$ & $C_2$ & $0.047\mu F$ \\\hline
        $R_4$ & $10k\Omega$ & $C_{fb}$ & $0.047\mu F$ \\\hline
        $R_5$ & $1k\Omega$ & $C_{ff}$ & $0.047\mu F$ \\\hline
        $R_6$ & $200k\Omega$ & $C_{LP}$ & $0.47\mu F$ \\\hline
        $R_7$ & $82k\Omega$ & $Vcc$ & $10V$ \\\hline
        $R_8$ & $1\Omega$ & $V_A$ & $2V$ \\\hline
        $R_9$ & $1M\Omega$ & $V_B$ & $1V$ \\\hline
    \end{tabular}
    \end{center}
\end{table}

In future work, the tuning of the feedforward gain, and other parameters of the neurons and their connections, could be done by some sort of adaptation or learning mechanism within the system architecture. This would make the system more robust to disturbances and more representative of biology, where evidence of learning is also found in homeostatic processes \cite{learning}. Another improvement is to make the setpoint not system inherent by using a temperature insensitive neuron as a reference signal, see \cite{setpoint}.




\bibliography{IEEEabrv,References}

\begin{thebibliography}{10}

\bibitem{animaltemp}
K.~Kanosue, L.~Crawshaw, K.~Nagashima, and T.~Yoda, ``Concepts to utilize in
  describing thermoregulation and neurophysiological evidence for how the
  system works,'' {\em European journal of applied physiology}, vol.~109,
  pp.~5--11, 10 2009.

\bibitem{thermomechanisms}
S.~Morrison and K.~Nakamura, ``Central mechanisms for thermoregulation,'' {\em
  An. Rev. Physio.}, vol.~81, no.~Volume 81, pp.~285--308, 2019.

\bibitem{original}
P.~Feketa, T.~Birkoben, M.~Noll, A.~Schaum, T.~Meurer, and H.~Kohlstedt,
  ``Artificial homeostatic temperature regulation via bio-inspired feedback
  mechanisms,'' {\em Scient. Rep.}, vol.~13, no.~1, p.~5003, 2023.

\bibitem{hybridsystems}
R.~Goebel, R.~G. Sanfelice, and A.~R. Teel, {\em Hybrid Dynamical Systems:
  Modeling, Stability, and Robustness}.
\newblock Princeton University Press, 2012.

\bibitem{lapicque1907recherches}
L.~Lapicque, ``Recherches quantitatives sur l’excitation \'electrique des
  nerfs trait\'ee comme une polarization,'' {\em J. Physiol. Pathol. Gen.
  9:620-635}, 1907.

\bibitem{abbott1999lapicque}
L.~F. Abbott, ``Lapicque’s introduction of the integrate-and-fire model
  neuron (1907),'' {\em Brain res. bul.}, vol.~50, no.~5-6, pp.~303--304, 1999.

\bibitem{izhikevich2010hybrid}
E.~M. Izhikevich, ``Hybrid spiking models,'' {\em Philosophical Transactions of
  the Royal Society A: Mathematical, Physical and Engineering Sciences},
  vol.~368, no.~1930, pp.~5061--5070, 2010.

\bibitem{NTCLE100E3}
Vishay, ``Ntcle100e3 product information,'' 2025.
\newblock \url{https://www.vishay.com/en/product/29049/} (visited 2025-03-18).

\bibitem{Si7137DP}
Vishay, ``Si7137dp mosfet product information,'' 2025.
\newblock \url{https://www.vishay.com/en/product/69063/} (visited 2025-03-11).

\bibitem{hybridtoolbox}
R.~Sanfelice, ``Hybrid equations toolbox,'' 2022.
\newblock
  \url{https://nl.mathworks.com/matlabcentral/fileexchange/41372-hybrid-equations-toolbox}
  (visited 2025-03-21).

\bibitem{sepulchre1}
R.~Sepulchre, ``Spiking control systems,'' {\em Proceedings of the IEEE},
  vol.~110, no.~5, pp.~577--589, 2022.

\bibitem{petri1}
E.~Petri, K.~J.~A. Scheres, E.~Steur, and W.~P. M.~H. Heemels, ``Analysis of a
  simple neuromorphic controller for linear systems: A hybrid systems
  perspective,'' in {\em IEEE Conference on Decision and Control, Milan,
  Italy}, pp.~8578--8583, 2024.

\bibitem{sepulchre2}
R.~Schmetterling, F.~Forni, A.~Franci, and R.~Sepulchre, ``Neuromorphic control
  of a pendulum,'' {\em IEEE Control Systems Letters}, vol.~8, pp.~1235--1240,
  2024.

\bibitem{learning}
D.~Ramsay, S, and S.~Woods, ``Physiological regulation: How it really works,''
  {\em Cell Metabolism}, vol.~24, no.~3, pp.~361--364, 2016.

\bibitem{setpoint}
J.~A. Boulant, ``Neuronal basis of {H}ammel's model for set-point
  thermoregulation,'' {\em Journal of Applied Physiology}, vol.~100, no.~4,
  pp.~1347--1354, 2006.

\end{thebibliography}


\end{document}